\documentclass[graybox]{svmult}

\usepackage{mathptmx}       
\usepackage{helvet}         
\usepackage{courier}        
\usepackage{type1cm}        
\usepackage{makeidx}         
\usepackage{graphicx}        
\usepackage{multicol}        
\usepackage[bottom]{footmisc}

\newcommand\rf[1]{(\ref{eq:#1})}
\newcommand\lab[1]{\label{eq:#1}}
\newcommand\nonu{\nonumber}
\newcommand\br{\begin{eqnarray}}
\newcommand\er{\end{eqnarray}}
\newcommand\be{\begin{equation}}
\newcommand\ee{\end{equation}}

\newcommand\foot[1]{\footnotemark\footnotetext{#1}}
\newcommand\lb{\lbrack}
\newcommand\rb{\rbrack}

\newcommand\llb{\left\lbrack}
\newcommand\rrb{\right\rbrack}

\renewcommand\({\left(}
\renewcommand\){\right)}
\newcommand\bv{\bigm\vert}               

\newcommand\bc{\begin{center}}
\newcommand\ec{\end{center}}

\newcommand\partder[2]{\frac{{\partial {#1}}}{{\partial {#2}}}}

\renewcommand\d{\delta}

\newcommand\g{\gamma}

\newcommand\h{\frac{1}{2}}

\renewcommand\l{\lambda}

\newcommand\m{\mu}
\newcommand\n{\nu}

\newcommand\vp{\varphi}

\newcommand\pa{\partial}

\newcommand\s{\sigma}

\renewcommand\t{\tau}
\renewcommand\th{\theta}

\newcommand\wti{\widetilde}

\newcommand\cA{{\mathcal A}}

\newcommand\cE{{\mathcal E}}

\newcommand\cJ{{\mathcal J}}

\newcommand\cW{{\mathcal W}}

\newcommand{\ct}[1]{\cite{#1}}
\newcommand{\bib}[1]{\bibitem{#1}}

\newcommand\PRD[3]{\textsl{Phys. Rev.} \textbf{D#1} (#2) #3}

\newcommand\PLB[3]{\textsl{Phys. Lett.} \textbf{#1B} (#2) #3}

\newcommand\MPLA[3]{\textsl{Mod. Phys. Lett.} \textbf{A#1} (#2) #3}

\newcommand\xdot{\stackrel{.}{x}}

\begin{document}

\title*{Kruskal-Penrose Formalism for Lightlike Thin-Shell Wormholes}
\author{Eduardo Guendelman, Emil Nissimov, Svetlana Pacheva and Michail Stoilov}
\institute{Eduardo Guendelman \at Department of Physics, Ben-Gurion University of
the Negev, Beer-Sheva, Israel \email{guendel@bgu.ac.il}
\and Emil Nissimov, Svetlana Pacheva and Michail Stoilov
\at Institute for Nuclear Research and Nuclear Energy,
Bulgarian Academy of Sciences, Sofia, Bulgaria \\
\email{nissimov@inrne.bas.bg, svetlana@inrne.bas.bg, mstoilov@inrne.bas.bg}}
%
%
\maketitle

\abstract{The original formulation of the ``Einstein-Rosen bridge'' in the
classic paper of Einstein and Rosen (1935) is historically the first 
example of a {\em static} spherically-symmetric wormhole solution. It is 
{\em not equivalent} to the concept of the {\em dynamical} and 
{\em non-traversable} Schwarzschild wormhole, also called ``Einstein-Rosen bridge''
in modern textbooks on general relativity. In previous papers of ours we
have provided a mathematically correct treatment of the original 
``Einstein-Rosen bridge'' as a {\em traversable} wormhole by showing that it
requires the presence of a special kind of ``exotic matter'' located on the
wormhole throat -- a lightlike brane (the latter was overlooked in the
original 1935 paper). In the present note we continue our thorough study of the
original ``Einstein-Rosen bridge'' as a simplest example of
a lightlike thin-shell wormhole by explicitly deriving its description in
terms of the Kruskal-Penrose formalism for maximal analytic extension of the
underlying wormhole spacetime manifold. Further, we generalize the
Kruskal-Penrose description to the case of more complicated lightlike
thin-shell wormholes with two throats exhibiting a remarkable property of
QCD-like charge confinement. 
}

\section{Introduction}
\label{intro}

The principal object of study in the present note is the class of static
spherically symmetric {\em lightlike thin-shell} wormhole solutions in
general relativity, \textsl{i.e.}, spacetimes with wormhole geometries and ``throats''
being lightlike (``null'') hypersurfaces (for the importance and impact of
lightlike hypersurfaces, see
Refs.\ct{barrabes-israel-91,barrabes-hogan,barrabes-israel-05}). 
The explicit construction of lightlike thin-shell wormholes based on a
self-consistent Lagrangian action formalism for the underlying lightlike
branes occupying the wormhole ``throats'' and serving as material (and
electrical charge) sources for the gravity to generate the wormhole spacetime geometry was
given in a series of previous papers \ct{kerr-wormhole}-\ct{BR-kink}
\foot{For the general construction of {\em timelike} thin-shell wormholes,
see the book \ct{visser-book}}.

The celebrated ``Einstein-Rosen bridge'', originally formulated in the
classic paper \ct{ER-1935}, is historically the first and simplest 
example of a {\em static} spherically-symmetric wormhole solution -- it is a
4-dimensional spacetime manifold consisting of two identical copies of the
exterior Schwarzschild spacetime region matched (glued together) along their
common horizon. 

Let us immediately emphasize that the original construction
in \ct{ER-1935} of the ``Einstein-Rosen bridge'' is {\em not equivalent} to the 
notion of the {\em dynamical} Schwarzschild wormhole, also called 
``Einstein-Rosen bridge'' in several standard textbooks (\textsl{e.g.}
Ref.\ct{MTW}), which employs the formalism of Kruskal-Szekeres maximal
analytic extension of Schwarzschild black hole spacetime geometry. Namely, 
the two regions in Kruskal-Szekeres manifold corresponding to the outer 
Schwarzschild spacetime region beyond the horizon ($r>2m$) and labeled $(I)$ and $(III)$ 
in Ref.\ct{MTW} are generally {\em disconnected} and share only a two-sphere
(the angular part) as a common border ($U=0, V=0$ in Kruskal-Szekeres coordinates), 
whereas in the original Einstein-Rosen ``bridge'' construction the boundary between 
the two identical copies of the outer Schwarzschild space-time region ($r>2m$) is a 
three-dimensional lightlike hypersurface ($r=2m)$. Physically, the most
significant difference is that the ``textbook'' version of the ``Einstein-Rosen bridge''
(Schwarzschild wormhole) is {\em non-traversable}, \textsl{i.e.}, there are
{\em no} timelike or lightlike geodesics connecting points belonging to the
two separate outer Schwarzschild regions $(I)$ and $(III)$. This is in sharp
contrast w.r.t. the original Einstein-Rosen bridge (within its consistent
formulation as a lightlike thin-shell wormhole \ct{ER-bridge}), which is a 
{\em traversable wormhole} (see also Section 3 below).

However, as explicitly demonstrated in Refs.\ct{ER-bridge,rotating-WH}, the
originally proposed in \ct{ER-1935} Einstein-Rosen ``bridge'' wormhole solution 
{\em does not} satisfy the vacuum Einstein equations
at the wormhole ``throat''. The mathematically consistent formulation of the original
Einstein-Rosen ``bridge'' requires solving Einstein equations of bulk $D=4$
gravity coupled to a lightlike brane with a well-defined world-volume
action \ct{will-prd}-\ct{varna-07}. The lightlike brane
locates itself automatically on the wormhole throat glueing together the two
``universes'' - two identical copies of the external spacetime region of a 
Schwarzschild black hole matched at their common horizon, with a special relation 
between the (negative) brane tension and the Schwarzschild mass parameter. 
This is briefly reviewed in Section 2.

Traversability of the correctly formulated Einstein-Rosen bridge as a
lightlike thin-shell wormhole is explicitly demonstrated in Section 3 in the sense of
passing through the wormhole throat from the ``left'' to the ``right'' universe 
within {\em finite proper time} of a travelling observer.

In Sections 4 we explicitly construct the Kruskal-Penrose maximal
analytic extension of the proper Einstein-Rosen bridge wormhole manifold. In
particular, the pertinent Kruskal-Penrose manifold involves a special identification 
of the future horizon of the ``right'' universe with the past horizon of the ``left'' 
universe, which is the mathematical manifestation of the wormhole traversability.

In Section 5 we extend our construction of Kruskal-Penrose maximal analytic extension
of the total wormhole manifold to the case of a physically interesting wormhole 
solution with two ``throats'' which exhibits a remarkable property of charge 
and electric flux confinement \ct{hide-confine} resembling the quark 
confinement property of quantum chromodynamics.

Section 6 contains our concluding remarks.


\section{Einstein-Rosen Bridge as Lightlike Thin-Shell Wormhole}
\label{ER-bridge+LL-brane}

The Schwarzschild spacetime metric is the simplest static spherically symmetric 
black hole metric, written in standard coordinates $(t,r,\th,\vp)$
(textsl{e.g. \ct{MTW}):
\be
ds^2 = - A(r) dt^2 + \frac{1}{A(r)} dr^2 + r^2 \( d\th^2 + \sin^2\th d\vp^2\) 
\quad ,\quad A(r) = 1 - \frac{r_0}{r} \; ,
\lab{Schw-metric}
\ee
where $r_0 \equiv 2m$ ($m$ -- black hole mass parameter):

\begin{itemize}
\item
$r>r_0$ defines the exterior spacetime region;  $r<r_0$ is the black hole region;
\item
$r_0$ is the horizon radius, where $A(r_0) = 0$ ($r=r_0$ is a
non-physical coordinate singularity of the metric \rf{Schw-metric}, unlike
the physical spacetime singularity at $r=0$).
\end{itemize}

In constructing the maximal analytic extension of the Schwarzschild spacetime 
geometry -- the {\em Kruskal-Szekeres} coordinate chart -- essential intermediate
use is made of the so called ``tortoise'' coordinate $r^{*}$ 
(for light rays $t\pm r^{*}={\rm const}$):
\be
\frac{dr^{*}}{dr}=\frac{1}{A(r)} \quad \longrightarrow \quad
r^{*} = r + r_0 \ln |r-r_0| \; .
\lab{tortoise}
\ee

The Kruskal-Szekeres (``light-cone'') coordinates $(v,w)$ 
are defined as follows (\textsl{e.g.} \ct{MTW}):
\be
v = \pm \frac{1}{\sqrt{2k_h}} e^{k_h \bigl(t+r^{*}\bigr)} \quad ,\quad
w = \mp \frac{1}{\sqrt{2k_h}} e^{-k_h \bigl(t-r^{*}\bigr)} 
\lab{KS-coord}
\ee
with all combinations of the overall signs,
where $k_h = \h \pa_r A(r)\bv_{r=r_0} = \frac{1}{2r_0}$ is the so called
``surface gravity'' (related to the Hawking temperature as
$\frac{k_h}{2\pi}=k_B T_{\rm hawking}$). Eqs.\rf{KS-coord} are equivalent to:
\be
\mp vw = \frac{1}{k_h} e^{2 k_h r^{*}} \quad ,\quad \mp\frac{v}{w} = e^{2 k_h t} \; ,
\lab{KS-coord-1}
\ee
wherefrom $t$ and $r^{*}$ are determined as functions of $vw$. 

Depending on the combination of the overall signs Eqs.\rf{KS-coord} define a
doubling the regions of the standard Schwarzschild geometry \ct{MTW}:

(i) $(+,-)$ -- exterior Schwarzschild region $r>r_0$ (region $I$); 

(ii) $(+,+)$ -- black hole $r<r_0$ (region $II$); 

(iii) $(-,+)$ -- second copy of exterior Schwarzschild region $r>r_0$ (region $III$); 

(iv) $(-,-)$ -- ``white'' hole region $r<r_0$ (region $IV$).

The metric \rf{Schw-metric} becomes:
\be
ds^2 = {\wti A}(vw) dv dw +
r^2(vw) \( d\th^2 + \sin^2\th d\vp^2\) \quad,
\quad {\wti A}(vw) \equiv \frac{A\bigl(r(vw)\bigr)}{k_h^2 vw} \; ,
\lab{KS-metric}
\ee
so that now there is no coordinate singularity on the horizon ($v=0$ or $w=0$)
upon using Eq.\rf{tortoise}: ${\wti A}(0) = -4$ .

In the classic paper \ct{ER-1935} Einstein and Rosen introduced in \rf{Schw-metric}
a new radial-like coordinate $u$ via $r = r_0 + u^2$ and let 
$u \in (-\infty, +\infty)$:
\be
ds^2 = - \frac{u^2}{u^2 + r_0} dt^2 + 4 (u^2 + r_0)du^2 +
(u^2 + r_0)^2 \( d\th^2 + \sin^2 \th \,d\vp^2\) \; .
\lab{E-R-metric}
\ee
Thus, \rf{E-R-metric} describes two identical copies of the exterior
Schwarzschild spacetime region ($r> r_0$) for $u>0$ and $u<0$, which 
are formally glued together at the horizon $u=0$.

Unfortunately, there are serious problems with \rf{E-R-metric}:
\begin{itemize}
\item
The Einstein-Rosen metric \rf{E-R-metric} has coordinate singularity at $u=0$: 
$\det\Vert g_{\m\n}\Vert_{u=0} = 0$.
\item
More seriously, the Einstein equations for \rf{E-R-metric} acquire an 
ill-defined non-vanishing ``matter'' stress-energy
tensor term on the r.h.s., which was overlooked in the original 1935 paper!
\end{itemize}

Indeed, as explained in \ct{ER-bridge}, from Levi-Civita identity 
$R^0_0 = - \frac{1}{\sqrt{-g_{00}}} \nabla^2_{(3)} \(\sqrt{-g_{00}}\)$ we
deduce that \rf{E-R-metric} solves vacuum Einstein eq. $R^0_0 = 0$ for all
$u\neq 0$. However, since $\sqrt{-g_{00}} \sim |u|$ as $u \to 0$ and since
$\frac{\pa^2}{{\pa u}^2} |u| = 2 \d (u)$, Levi-Civita identity tells us that:
\be
R^0_0 \sim \frac{1}{|u|} \d (u) \sim \d (u^2) \; ,
\lab{ricci-delta}
\ee
and similarly for the scalar curvature $R \sim \frac{1}{|u|} \d (u) \sim \d (u^2)$.

In \ct{ER-bridge} we proposed a correct reformulation of the original
Einstein-Rosen bridge as a mathematically consistent traversable 
lightlike thin-shell wormhole introducing a different radial-like coordinate 
$\eta \in (-\infty, +\infty)$, by
substituting $r = r_0 + |\eta|$ in \rf{Schw-metric}:
\be
ds^2 = - \frac{|\eta|}{|\eta| + r_0} dt^2 + \frac{|\eta| + r_0}{|\eta|}d\eta^2 +
(|\eta| + r_0)^2 \( d\th^2 + \sin^2 \th \,d\vp^2\) \; .
\lab{our-ER-metric}
\ee

Eq.\rf{our-ER-metric} is the correct spacetime metric for the original
Einstein-Rosen bridge:

\begin{itemize}
\item
Eq.\rf{our-ER-metric} describes two ``universes'' -- two identical copies of
the exterior Schwarzschild spacetime region for $\eta >0$ and $\eta <0$.
\item
Both ``universes'' are correctly glued together at their common horizon $\eta=0$. 
Namely, the metric \rf{our-ER-metric} solves Einstein equations:
\be
R_{\m\n} - \h g_{\m\n} R = 8\pi T^{(brane)}_{\m\n} \; ,
\lab{Einstein-eqs+LL}
\ee
where on the r.h.s. $T^{(brane)}_{\m\n} = S_{\m\n} \d (\eta)$ is
the energy-momentum tensor of a special kind of {\em lightlike brane}
located on the common horizon $\eta=0$ -- the wormhole ``throat''.
\item
The lightlike analogues of W.Israel's junction conditions on the 
wormhole ``throat'' are satisfied \ct{ER-bridge,rotating-WH}.
\item
The resulting lightlike thin-shell wormhole is traversable (see Section 3 below).
\end{itemize}

The energy-momentum tensor of lightlike branes $T^{(brane)}_{\m\n}$ is 
self-consistently derived as 
$T^{(brane)}_{\m\n} = - \frac{2}{\sqrt{-g}} \frac{\d S_{\rm LL}}{\d g^{\m\n}}$
from the following manifestly reparametrization invariant world-volume Polyakov-type 
lightlike brane action (written for arbitrary $D=(p+1)+1$ embedding
spacetime dimension and $(p+1)$-dimensional brane world-volume):
\br
S_{\rm LL} = - \h \int d^{p+1} \s\, T b_0^{\frac{p-1}{2}}\sqrt{-\g}
\llb \g^{ab} {\bar g}_{ab} - b_0 (p-1)\rrb \; ,
\lab{LL-action} \\
{\bar g}_{ab} \equiv 
g_{ab} - \frac{1}{T^2} \bigl(\pa_a u + q\cA_a\bigr)\,\bigl(\pa_b u + q\cA_b\bigr) 
\;\; ,\;\; \cA_a \equiv \pa_a X^\m A_\m \; .
\lab{ind-metric-ext}
\er
Here and below the following notations are used:
\begin{itemize}
\item
$\g_{ab}$ is the {\em intrinsic} Riemannian metric on the world-volume with
$\g = \det \Vert\g_{ab}\Vert$;
$b_0$ is a positive constant measuring the world-volume ``cosmological constant'';
$(\s)\equiv \(\s^a \)$ with $a=0,1,\ldots ,p$ ; $\pa_a \equiv \partder{}{\s^a}$.
\item
$X^\m (\s)$ are the $p$-brane embedding coordinates in the bulk
$D$-dimensional spacetime with Riemannian metric 
$g_{\m\n}(x)$ ($\m,\n = 0,1,\ldots ,D-1$). 
$A_\m$ is a spacetime electromagnetic field (absent in the present case).
\item
$g_{ab} \equiv \pa_a X^{\m} g_{\m\n}(X) \pa_b X^{\n}$ 
is the {\em induced} metric on the world-volume 
which becomes {\em singular} on-shell --  
manifestation of the lightlike nature of the brane.
\item
$u$ is auxiliary world-volume scalar field defining the lightlike direction
of the induced metric 
and it is a non-propagating degree of freedom. 
\item
$T$ is {\em dynamical (variable)} brane tension (also a non-propagating
degree of freedom).
\item
Coupling parameter $q$ is the surface charge density of the LL-brane ($q=0$
in the present case).
\end{itemize}

The Einstein Eqs.\rf{Einstein-eqs+LL} imply the following relation between 
the lightlike brane parameters and the Einstein-Rosen bridge ``mass'' ($r_0 = 2m$):
\be
-T=\frac{1}{8\pi m} \;\; ,\;\; b_0 = \frac{1}{4} \; ,
\lab{rel-param}
\ee
\textsl{i.e.}, the lightlike brane dynamical tension $T$ becomes {\em negative}
on-shell -- manifestation of ``exotic matter'' nature.



\section{Einstein-Rosen Bridge as Traversable Wormhole}
\label{traversability}

As already noted in \ct{ER-bridge,rotating-WH} traversability of the
original Einstein-Rosen bridge is a particular manifestation of the
traversability of lightlike ``thin-shell'' wormholes \foot{Subsequently,
traversability of the Einstein-Rosen bridge has been studied using
Kruskal-Szekeres coordinates for the Schwarzschild black hole \ct{poplawski}, 
or the 1935 Einstein-Rosen coordinate chart \rf{E-R-metric} \ct{katanaev}.}.
Here for completeness we will present the explicit details of the
traversability within the proper Einstein-Rosen bridge wormhole coordinate
chart \rf{our-ER-metric} which are needed for the construction of the
pertinent Kruskal-Penrose diagram in Section 4.

The motion of test-particle (``observer'') of mass $m_0$ in a gravitational background 
is given by the reparametrization-invariant world-line action:
\be
S_{\rm particle} = 
\h \int d\l \llb \frac{1}{e} g_{\m\n} \xdot^\m \xdot^\n - e m_0^2\rrb \; ,
\lab{particle-action}
\ee
where $\xdot^\m \equiv \frac{dx^\m}{d\l}$, $e$ is the world-line ``einbein'' and
in the present case $(x^\m) = (t,\eta,\th,\vp)$.

For a static spherically symmetric background such as \rf{our-ER-metric} 
there are conserved Noether ``charges'' -- energy $\cE$ and angular momentum $\cJ$. 
In what follows we will consider purely ``radial'' motion ($\cJ=0$) so, upon taking 
into account the ``mass-shell'' constraint (the equation of motion w.r.t. $e$) 
and introducing the world-line proper-time parameter $\t$ ($\frac{d\t}{d\l}=e m_0$), 
the timelike geodesic equations (world-lines of massive point particles) read: 
\be
\Bigl(\frac{d\eta}{d\t}\Bigr)^2 = \frac{\cE^2}{m_0^2} - A(\eta) \;\; ,\;\;
\frac{dt}{d\t}= \frac{\cE}{m_0 A(\eta)} \;\; ,\;\; 
A(\eta) \equiv \frac{|\eta|}{|\eta| + r_0} \; .
\lab{radial-eqs}
\ee
where $A(\eta)$ is the ``$-g_{00}$'' component of the proper Einstein-Rosen 
bridge metric \rf{our-ER-metric}.

For a test-particle starting for $\t=0$ at initial position in ``our''
(right) universe $\eta_0 = \eta (0)\; , t_0 = t(0)$ and {\em infalling}
towards the ``throat'' the solutions of Eqs.\rf{radial-eqs} read:
\br
\frac{\cE}{2k_h m_0} \int^{2k_h \eta_0}_{2k_h \eta (\t)} dy
\sqrt{(1+|y|)\Bigl\lb(1+\bigl(1-\frac{m_0^2}{\cE^2}\bigr)|y|\Bigr\rb^{-1}}
= \t \; ,
\lab{eta-eq}\\
\frac{1}{2k_h} \int^{2k_h \eta_0}_{2k_h \eta (\t)} dy \frac{1}{|y|} 
\sqrt{(1+|y|)\Bigl\lb(1+\bigl(1-\frac{m_0^2}{\cE^2}\bigr)|y|\Bigr\rb}
= t(\t)-t_0 \; .
\lab{t-eq}
\er

\begin{itemize}
\item
Eq.\rf{eta-eq} shows that the particle will cross the wormhole ``throat'' 
($\eta=0$) for a finite proper-time $\t_0 >0$:
\be
\t_0 = \frac{\cE}{2k_h m_0} \int^{2k_h \eta_0}_{0} dy
\sqrt{(1+|y|)\Bigl\lb(1+\bigl(1-\frac{m_0^2}{\cE^2}\bigr)|y|\Bigr\rb^{-1}}\; .
\lab{eta-eq-0}
\ee
\item
It will continue into the second (left) universe and reach any point 
$\eta_1 = \eta (\t_1) <0$ within another \textbf{finite} proper-time $\t_1 >\t_0$.
\item
On the other hand, from \rf{t-eq} it follows that $t(\t_0 - 0)=+\infty$, 
\textsl{i.e.}, from the point of view of a static observer in ``our'' (right) 
universe it will take infinite ``laboratory'' time for the particle to reach the 
``throat'' -- the latter appears to the static observer as a future black hole horizon.
\item
Eq.\rf{t-eq} also implies $t(\t_0 + 0)= -\infty$, which means that from the point 
of view of a static observer in the second (left) universe, upon crossing the
``throat'', the particle starts its motion in the second (left) universe
from infinite past, so that it will take an infinite amount of ``laboratory''
time to reach the point $\eta_1 <0$ -- \textsl{i.e.} the ``throat'' now
appears as a past black hole horizon.
\end{itemize}

In analogy with the usual ``tortoise'' coordinate $r^{*}$ for the
Schwarzschild black hole geometry \rf{tortoise} let us now introduce
Einstein-Rosen bridge ``tortoise'' coordinate $\eta^{*}$ 
(recall $r_0 = \frac{1}{2k_h}$):
\be
\frac{d\eta^{*}}{d\eta} = \frac{|\eta|+r_0}{|\eta|} \quad \longrightarrow
\quad \eta^{*} = \eta + {\rm sign} (\eta) r_0 \ln |\eta| \; .
\lab{ER-tortoise}
\ee
Let us note here an important difference in the behavior of the
``tortoise'' coordinates $r^{*}$ \rf{tortoise} and $\eta^{*}$
\rf{ER-tortoise} in the vicinity of the horizon. Namely:
\be
r^{*} \to -\infty \quad {\rm for} \;\; r \to r_0 \pm 0 \; ,
\lab{tortoise-r}
\ee
\textsl{i.e.}, when $r$ approaches the horizon either from above or from below,
whereas when $\eta$ approaches the horizon from above or from below:
\be
\eta^{*} \to \mp \infty \quad {\rm for} \;\; \eta \to \pm 0 \; .
\lab{tortoise-eta}
\ee

For infalling/outgoing massless particles (light rays) 
Eqs.\rf{eta-eq}-\rf{ER-tortoise} imply:
\be
t\pm \eta^{*} = {\rm const} \; .
\lab{t-plus-eta-star-1}
\ee
For infalling massive particles towards the ``throat'' ($\eta=0$) starting at
$\eta^{+}_0 >0$ in ``our'' (right) universe and crossing into the second 
(left) universe, or starting in the second (left)
universe at some $\eta^{-}_0 <0$ and crossing into the ``our'' (right) universe, 
we have correspondingly (replacing $\t$-dependence with functional
dependence w.r.t. $\eta$ using first Eq.\rf{radial-eqs}):
\be
\bigl\lb t\pm\eta^{*}\bigr\rb (\eta) = \frac{\pm 1}{2k_h}
\int^{2k_h \eta^{\pm}_0}_{2k_h \eta} dy \Bigl( 1+ \frac{1}{|y|}\Bigr)
\llb \sqrt{(1+|y|)\Bigl\lb(1+\bigl(1-\frac{m_0^2}{\cE^2}\bigr)|y|\Bigr\rb^{-1}}
-1 \rrb \; .
\lab{t-plus-eta-star-2}
\ee


\section{Kruskal-Penrose Diagram for Einstein-Rosen Bridge}
\label{Kruskal-Penrose-ER-bridge}

We now define the maximal analytic extension of original Einstein-Rosen wormhole
geometry \rf{our-ER-metric} via introducing Kruskal-like coordinates $(v,w)$
as follows:
\be
v = \pm \frac{1}{\sqrt{2k_h}} e^{\pm k_h (t+\eta^{*})} \;\; , \;\;
w = \mp \frac{1}{\sqrt{2k_h}} e^{\mp k_h (t-\eta^{*})} \; ,
\lab{region-1-2}
\ee
implying:
\be
-vw = \frac{1}{2k_h} e^{\pm 2k_h \eta^{*}} \;\; ,\;\;
-\frac{v}{w} = e^{\pm 2k_h t} \; .
\lab{region-1-2-a}
\ee
Here and below $\eta^{*}$ is given by \rf{ER-tortoise}.

\begin{itemize}
\item
The upper signs in \rf{region-1-2}-\rf{region-1-2-a} correspond to region $I$
$(v>0,w<0)$ describing ``our'' (right) universe $\eta>0$.
\item
The lower signs in \rf{region-1-2}-\rf{region-1-2-a} correspond to region $II$ 
$(v<0,w>0)$ describing the second (left) universe $\eta<0$.
\end{itemize}

The metric \rf{our-ER-metric} of Einstein-Rosen bridge in the Kruskal-like
coordinates \rf{region-1-2} reads:
\br
ds^2 = {\wti A}(vw) dv dw + {\wti r}^2 (vw) \bigl( d\th^2 + \sin^2 \th d\vp^2\Bigr) 
\; ,
\lab{ER-bridge-Kruskal} \\
{\wti r} (vw) = r_0 + |\eta (vw)| \quad (r_0 \equiv \frac{1}{2k_h}) \; , 
\nonu \\
{\wti A}(vw) = \frac{A\bigl(\eta (vw)\Bigr)}{k_h^2 vw} = 
- \frac{4 e^{-2k_h |\eta (vw)}|}{1 + 2k_h |\eta (vw)|} \; ,
\lab{ER-bridge-A}
\er
where $\eta (vw)$ is determined from \rf{region-1-2-a} and \rf{ER-tortoise} as:
\be
-vw = \frac{|\eta|}{2k_h} e^{2k_h |\eta|} \quad \longrightarrow \quad
|\eta (vw)| = \frac{1}{2k_h} \cW (-4k_h^2 vw) \; ,
\lab{vw-eq}
\ee
$\cW (z)$ being the Lambert (product-logarithm) function ($z=\cW (z) e^{\cW (z)}$).

Using the explicit expression \rf{ER-tortoise} for $\eta^{*}$ in 
\rf{region-1-2-a} we find for the metric \rf{ER-bridge-Kruskal}-\rf{ER-bridge-A}:
\begin{itemize}
\item
``Throats'' (horizons) -- at $v=0$ or $w=0$;
\item
In region $I$ the ``throat'' $(v>0,w=0)$ is a future horizon
$(\eta=0\, ,\, t\to +\infty)$, whereas the ``throat'' $(v=0,w<0)$ is a past horizon
$(\eta=0\, ,\, t\to -\infty)$.
\item
In region $II$ the ``throat'' $(v=0,w>0)$ is a future horizon
$(\eta=0\, ,\, t\to +\infty)$, whereas the ``throat'' $(v<0,w=0)$ is a past horizon
$(\eta=0\, ,\, t\to -\infty)$.
\end{itemize}

It is customary to replace Kruskal-like coordinates $(v,w)$ \rf{region-1-2} with 
compactified Penrose-like coordinates $({\bar v},{\bar w})$:
\be
{\bar v} = \arctan (\sqrt{2k_h}\, v) \quad ,\quad
{\bar w} = \arctan (\sqrt{2k_h}\, w) \; ,
\lab{Penrose-coord}
\ee
mapping the various ``throats'' (horizons) and infinities to finite lines/points:

\begin{itemize}
\item
In region $I$: future horizon $(0<{\bar v}<\frac{\pi}{2}, {\bar w}=0)$; 
past horizon $({\bar v}=0, -\frac{\pi}{2}<{\bar w} <0)$.
\item
In region $II$: future horizon $({\bar v}=0, 0<{\bar w}<\frac{\pi}{2})$; 
past horizon $(-\frac{\pi}{2}<{\bar v}<0, {\bar w}=0)$.
\item
$i_0$ -- spacelike infinity ($t={\rm fixed}, \eta \to \pm\infty$):\\
$i_0 = (\frac{\pi}{2},-\frac{\pi}{2})$ in region $I$;
$i_0 = (-\frac{\pi}{2},\frac{\pi}{2})$ in region $II$.
\item
$i_{\pm}$ -- future/past timelike infinity ($t \to \pm\infty, \eta={\rm fixed}$):\\
$i_{+}=(\frac{\pi}{2},0)$, $\, i_{-}=(0, -\frac{\pi}{2})$ in region $I$;
$i_{+}=(0,\frac{\pi}{2})$, $\, i_{-}=(-\frac{\pi}{2},0)$ in region $II$.
\item
$J_{+}$ -- future lightlike infinity ($t\to +\infty, \eta \to \pm\infty$, 
$\; t \mp \eta^{*} = {\rm fixed}$):\\
$J_{+} = ({\bar v}=\frac{\pi}{2}, -\frac{\pi}{2}<{\bar w}<0)$ in region $I$;\\
$J_{+} = (-\frac{\pi}{2}<{\bar v}<0,{\bar w}=\frac{\pi}{2})$ in region $II$.
\item
$J_{-}$ -- past lightlike infinity ($t\to -\infty, \eta \to \pm\infty$), 
$\; t \pm \eta^{*} = {\rm fixed}$):\\
$J_{-} = (0<{\bar v}<\frac{\pi}{2},{\bar w}= -\frac{\pi}{2})$ in region $I$:\\
$J_{-} = ({\bar v}= -\frac{\pi}{2},0<{\bar w}<\frac{\pi}{2})$ in region $II$.
\end{itemize}

For infalling light rays starting in region $I$ and crossing into
region $II$ we have the lightlike geodesic $t+\eta^{*}=c_1 \equiv {\rm const}$.
Thus, according to \rf{region-1-2} we must identify the crossing point $A$ on the
future horizon of region $I$ having Kruskal-like coordinates
$(v=\frac{1}{\sqrt{2k_h}}e^{k_h c_1},0)$ with the point $B$ on the past
horizon of region $II$ where the light rays enters into region $II$ whose 
Kruskal-like coordinates are $(v=-\frac{1}{\sqrt{2k_h}}e^{-k_h c_1},0)$.

Similarly, for infalling light rays starting in region $II$ and crossing into
region $I$ we have $t-\eta^{*}=c_2 \equiv {\rm const}$. Therefore,
the crossing point $C$ on the future horizon of region $II$ having 
Kruskal-like coordinates $(0,w=\frac{1}{\sqrt{2k_h}}e^{k_h c_2})$ must be identified
with the exit point $D$ 
$(0,w=-\frac{1}{\sqrt{2k_h}}e^{-k_h c_2})$ on the past horizon of region $I$.

Inserting Eqs.\rf{ER-tortoise}--\rf{t-plus-eta-star-2} into the definitions of 
Kruskal-like \rf{region-1-2} and Penrose-like \rf{Penrose-coord} coordinates
and taking into account the above identifications of horizons, 
we obtain the following visual representation of the Kruskal-Penrose diagram 
of the proper Einstein-Rosen bridge geometry \rf{our-ER-metric} 
as depicted in Fig.1:

\begin{figure}
\begin{center}
\includegraphics[width=11cm,keepaspectratio=true]{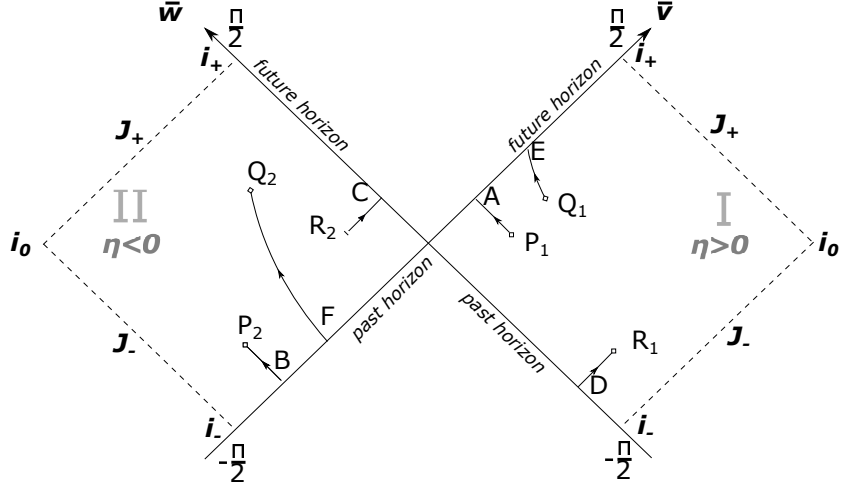}
\caption{Kruskal-Penrose diagram of the original Einstein-Rosen bridge}
\end{center}
\end{figure}

\begin{itemize}
\item
Future horizon in region $I$ is identified with past horizon in region $II$ as:
\be
({\bar v},0) \sim ({\bar v}-\frac{\pi}{2},0)\; .
\lab{ident-1}
\ee
Infalling light rays cross from region $I$ into region $II$ via paths 
$P_1 \to A\sim B \to P_2$ -- all the way within finite world-line time intervals
(the symbol $\sim$ means identification according to \rf{ident-1}).
Similarly, infalling massive particles cross from region $I$ into region $II$ 
via paths $ Q_1 \to E \sim F \to Q_2$ within finite proper-time interval.
\item
Future horizon in $II$ is identified with past horizon in $I$:
\be
(0,{\bar w}) \sim (0, {\bar w}-\frac{\pi}{2}) \; .
\lab{ident-2}
\ee
Infalling light rays cross from region $II$ into
region $I$ via paths $R_2 \to C\sim D \to R_1$ where $C\sim D$ is identified
according to \rf{ident-2}.
\end{itemize}


\section{Kruskal-Penrose Formalism for Two-Throat Lightlike Thin-Shell Wormhole}
\label{Kruskal-Penrose-charge-hiding-WH}

Now we will briefly discuss the extension of the construction of 
Kruskal-Penrose diagram for the
proper Einstein-Rosen bridge wormhole to the case of lightlike
``thin-shell'' wormholes with two throats. To this end we will consider the
physically interesting example of the charge-confining two-throat ``tube-like''
wormhole studied in \ct{hide-confine}. It is a solution of gravity interacting  
with a special non-linear gauge field system and both coupled to a pair of
oppositely charged lightlike branes (cf. Eqs.\rf{LL-action}-\rf{ind-metric-ext} above). 

The full wormhole spacetime consists of three ``universes'' glued pairwise
via the two oppositely charged lightlike branes located on their common horizons:

\begin{itemize}
\item
Region $I$: right-most non-compact electrically neutral ``universe'' -- exterior region
beyond the Schwarzschild horizon of a Schwarzschild-de Sitter black hole;
\item
Region $II$: middle ``tube-like'' ``universe'' of Levi-Civita-Bertotti-Robinson type 
\ct{LC-BR-1}-\ct{LC-BR-3} with finite radial-like spacial extend and 
compactified transverse spacial dimensions;
\item
Region $III$: left-most non-compact electrically neutral ``universe'' -- exterior region
beyond the Schwarzschild horizon of a Schwarzschild-de Sitter black hole,
mirror copy of the left-most ``universe''.
\item
Most remarkable property is that the whole electric flux generated by the two
oppositely charged lightlike branes sitting on the two ``throats'' 
is {\em completely confined} within the finite-spacial-size middle ``tube-like''
universe -- analog of QCD quark confinement!
\end{itemize}

For a visual representation, see Fig.2 \ct{hide-confine}.

\begin{figure}
\begin{center}
\includegraphics[width=11cm]{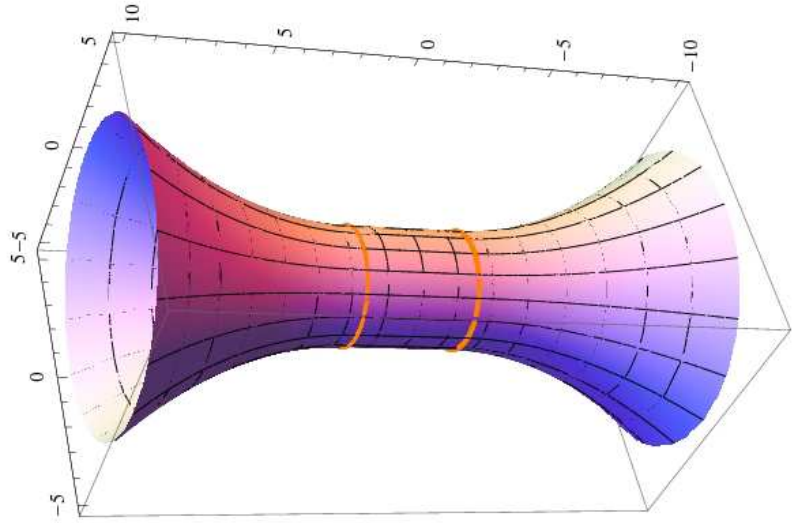}
\caption{Shape of $t=const$ and $\th=\frac{\pi}{2}$ slice of
charge-confining wormhole geometry. The whole electric flux is confined
within the middle cylindric ``tube'' (region $II$) connecting the two 
infinite ``funnels'' (region $I$ and region $III$). 
The rings on the edges of the ``tube'' depict the two oppositely
charged lightlike branes.}
\end{center}
\end{figure}

Generically, the metric of a spherically symmetric traversable lightlike thin-shell 
wormhole with two ``throats'' reads \ct{hide-confine} ($-\infty < \eta < \infty$):
\br
ds^2 = - A(\eta) dt^2 + \frac{d\eta^2}{A(\eta)} 
+ r^2 (\eta) \(d\th^2 + \sin^2\th d\vp^2\) \; , \phantom{aaaaaaa} 
\lab{2-throat-metric-0} \\
A(\eta_1) = 0\; ,\; A(\eta_2)=0 \; ,\;
a^{(1)}_{(\pm)}= \pm \partder{}{\eta} A \bv_{\eta_1 \pm 0} > 0 \; ,\;
a^{(2)}_{(\pm)}= \pm \partder{}{\eta} A \bv_{\eta_2 \pm 0} > 0 \; .
\lab{2-throat-metric}
\er
Accordingly, for the wormhole ``tortoise'' coordinate $\eta^{*}$ defined as in
first Eq.\rf{ER-tortoise} we have in the vicinity of the two horizons $\eta_{1,2}$:
\br
\eta^{*} = {\rm sign}(\eta -\eta_1) a^{(1)}_{(\pm)} \ln |\eta -\eta_1| +
O\bigl((\eta -\eta_1)^2\bigr) \; ,
\lab{2-throat-tortoise-1} \\
\eta^{*} = {\rm sign}(\eta -\eta_2) a^{(2)}_{(\pm)} \ln |\eta -\eta_2| +
O\bigl((\eta -\eta_2)^2\bigr) \; .
\lab{2-throat-tortoise-2}
\er

Now we can introduce the Kruskal-like and the compactified
Kruskal-Penrose coordinates $({\bar v},{\bar w})$ for the maximal analytic 
extension of the two-throat lightlike thin-shell wormhole generalizing formulas 
\rf{region-1-2} and \rf{Penrose-coord} as follows:

\begin{itemize}
\item
In region $I$ (right-most universe) -- $(+\infty >\eta > \eta_1)$:
\be
{\bar v}, {\bar w} = \pm \frac{\pi}{2\sqrt{a^{(1)}_{(-)}}} 
\pm \frac{1}{\sqrt{a^{(1)}_{(+)}}}
\arctan \Bigl( e^{\h a^{(1)}_{(+)} (\eta^{*} \pm t)}\Bigr)
\lab{region-1}
\ee
\item
In region $II$ (middle universe) -- $(\eta_1 >\eta > \eta_2)$; here 
$a^{(1)}_{(-)}=a^{(2)}_{(+)}$ which is satisfied in the case of the
charge-confining two-throat ``tube'' wormhole:
\be
{\bar v}, {\bar w} = \pm \frac{1}{\sqrt{a^{(1)}_{(-)}}}
\arctan \Bigl( e^{\h a^{(1)}_{(-)} (\eta^{*}\pm t)}\Bigr) \; .
\lab{region-2}
\ee
\item
In region $III$ (left-most universe) -- $(\eta_2 >\eta > -\infty)$:
\be
{\bar v}, {\bar w} = \mp \frac{\pi}{2\sqrt{a^{(2)}_{(-)}}} 
\pm \frac{1}{\sqrt{a^{(2)}_{(-)}}}
\arctan \Bigl( e^{\h a^{(2)}_{(-)} (\eta^{*}\pm t)}\Bigr) \; .
\lab{region-3}
\ee
\end{itemize}

The resulting Kruskal-Penrose diagram is depicted on Fig.3.

\vspace{-0.1in}
\begin{figure}
\begin{center}
\includegraphics[width=12cm,keepaspectratio=true]{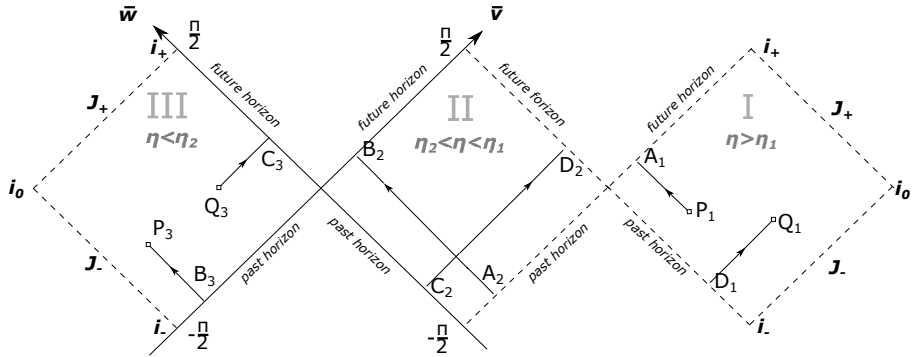}
\caption{Kruskal-Penrose diagram of ``charge-confining'' two-throat wormhole}
\end{center}
\end{figure}

In particular, infalling light ray starting in region $I$ arrives in region
$III$ within finite world-line time interval (``proper-time'' in the case of
massive particle) on the path $P_1 \to A_1 \sim A_2 \to B_2 \sim B_3 \to P_3$,
where the symbol $\sim$ indicates identification of the pertinent future and
past horizons of the ``glued'' together neighboring ``universes'' analogous to the
identification \rf{ident-1}, \rf{ident-2} in the simpler case of
Einstein-Rosen one-throat wormhole.

And similarly for an infalling light ray starting in region $III$ and arriving 
in region $I$ within finite world-line time interval on the path
$Q_3 \to C_3 \sim C_2 \to D_2 \sim D_1 \to Q_1$.

\section{Conclusions}
\label{conclude}

The mathematically correct reformulation \ct{ER-bridge} of original Einstein-Rosen 
``bridge'' construction, briefly reviewed in Section 2 above, shows that it is the 
simplest example in the class of static spherically symmetric \textsl{traversable}
lightlike ``thin-shell'' wormhole solutions in general relativity. The
consistency of Einstein-Rosen ``bridge'' as a traversable wormhole solution is 
guaranteed by the remarkable special properties of the world-volume dynamics
of the lightlike brane, which 
serves as an ``exotic'' thin-shell matter (and charge) source of gravity.

In the present note we have explicitly derived the Kruskal-like extension and
the associated Kruskal-Penrose diagram representation of the mathematically
correctly defined original Einstein-Rosen ``bridge'' \ct{ER-bridge}
with the following significant differences w.r.t. Kruskal-Penrose extension 
of the standard Schwarzschild black hole defining the corresponding ``textbook''
version of Einstein-Rosen ``bridge'' (the Schwarzschild wormhole) \ct{MTW}:

\begin{itemize}
\item
The pertinent Kruskal-Penrose diagram for the proper Einstein-Rosen bridge 
(Fig.1) has only two regions corresponding 
to ``our'' (right) and the second (left) ``universes'' unlike the four regions in the
standard Schwarzschild black hole case (no black/white hole regions).
\item
The proper original  Einstein-Rosen bridge is a
\textsl{traversable} static spherically symmetric wormhole unlike the
non-traversable non-static ``textbook'' version. Traversability is
equivalent to the pairwise specific identifications of future with past
horizons of the neighboring Kruskal regions.
\end{itemize}

We have also extended the Kruskal-Penrose diagram construction to the case
of lightlike ``thin-shell'' wormholes with two throats.


\begin{acknowledgement}
E.G., E.N. and S.P. gratefully acknowledge support of our collaboration through 
the academic exchange agreement between the Ben-Gurion University in Beer-Sheva,
Israel, and the Bulgarian Academy of Sciences. 
S.P. and E.N. have received partial support from European COST actions
MP-1210 and MP-1405, respectively.
E.N., S.P. and M.S. are also thankful to Bulgarian National Science Fund for
support via research grant DFNI-T02/6. 
\end{acknowledgement}
\end{document}